\documentclass[aps,prd,showpacs,nofootinbib,twocolumn,floatfix,superscriptaddress,preprintnumbers]{revtex4}
\usepackage{amsmath}
\usepackage{amssymb}
\usepackage{epsfig}
\usepackage{graphicx}
\usepackage{stmaryrd}
\usepackage{color}

\newcommand{\nn}{\nonumber}

\def\10{$SO(10)$}
\def\21{SU(2) $\otimes$ U(1) }

\def\422{$SU(4) \otimes SU(2) \otimes SU(2)$}
\def\321{SU(3) $\otimes$ SU(2) $\otimes$ U(1)}

\newcommand {\ignore}[1]{}

\def\lsim{\raise0.3ex\hbox{$\;<$\kern-0.75em\raise-1.1ex\hbox{$\sim\;$}}}
\def\gsim{\raise0.3ex\hbox{$\;>$\kern-0.75em\raise-1.1ex\hbox{$\sim\;$}}}
\def\vev#1{\left\langle #1\right\rangle}

\newcommand{\AddrIFIC}{%
  Institut de F\'{\i}sica Corpuscular --
  C.S.I.C./Universitat de Val{\`e}ncia \\
  Edificio Institutos de Paterna, Apt 22085, E--46071 Valencia, Spain}

 \newcommand{\ba}{\begin{array}}
\newcommand{\ea}{\end{array}}
\relax
\def\321{$SU(3)\times SU(2)\times U(1)$}

\newcommand{\Sol}  {\textrm{sol}}

\newcommand{\Atm}  {\textrm{atm}}

\newcommand{\Dms}  {\Delta m^2_\Sol}
\newcommand{\Dma}  {\Delta m^2_\Atm}

\begin{document}
\preprint{IFIC/09-32}
\renewcommand{\Huge}{\Large}
\renewcommand{\LARGE}{\Large}
\renewcommand{\Large}{\large}

\title{\bf Inverse tri-bimaximal type-III seesaw and lepton flavor violation}
\author{D.~Ibanez} \email{david.gil@ific.uv.es} \affiliation{\AddrIFIC}
\author{S.~Morisi} \email{morisi@ific.uv.es} \affiliation{\AddrIFIC}
\author{J.~W.~F.~Valle} \email{valle@ific.uv.es} \affiliation{\AddrIFIC}
\date{\today}

\begin{abstract}

  We present a type-III version of inverse seesaw or, equivalently an
  inverse version of type-III seesaw.
  Naturally small neutrino masses arise at low-scale from the exchange
  of neutral fermions transforming as hyperchargeless SU(2) triplets.
  In order to implement tri-bimaximal lepton mixing we supplement the
  minimal \321 gauge symmetry with an $A_4$-based flavor symmetry.
  Our scenario induces lepton flavour violating (LFV) $l_i\to l_j
  \bar{l}_k l_m$ decays that can proceed at the tree level, while
  radiative $l_i\to l_j \gamma $ decays and mu-e conversion in nuclei
  are also expected to be sizeable. LFV decays are related by the
  underlying flavor symmetry and the new fermions are also expected to
  be accessible for study at the Large Hadron Collider (LHC).

\end{abstract}
\maketitle

\section{Preliminaries}

Experiments~\cite{fukuda:2002pe,ahmad:2002jz,araki:2004mb,Kajita:2004ga,ahn:2002up}
have now confirmed that leptonic flavour is not conserved in nature:
the historical observation of neutrino oscillations has changed our
picture of fundamental physics. In contrast to the quark sector,
neutrino oscillations are characterized by two large mixing
angles~\cite{Maltoni:2004ei}. It is natural to expect that lepton
flavour violation (LFV) effects also take place among the electrically
charged partners of neutrinos under the weak interaction $SU(2)$.
The simplest and well-motivated way to induce neutrino LFV effects is
through the exchange of neutral leptons involved in generating
neutrino masses via various variants~\cite{Valle:2006vb} of the
simplest type-I
seesaw~\cite{gell-mann:1980vs,mohapatra:1981yp,schechter:1980gr,Lazarides:1980nt}.
The basic feature of such seesaw picture is that neutrino masses arise
only as a result of the exchange of heavy gauge singlet fermions
through
\begin{equation}
M_\nu=\left(
\begin{array}{cc}
0&M_D\\
M_D^T &M_R
\end{array}
\right),
\end{equation}
leading to an effective neutrino mass matrix
\begin{equation}\label{normal}
m_\nu=M_D^T M_R^{-1} M_D
\end{equation}
in the $(\nu,\nu^c)$ basis, where $\nu^c$ denote the heavy \321
singlet right-handed neutrino states which are sequentially added to
the Standard Model.
The smallness of neutrino mass follows naturally from the heaviness
of $\nu^c$.

\vskip5mm

As an alternative to the simplest type-I seesaw, it has long been
proposed that, thanks to the protecting $U(1)_L$ global lepton number
symmetry, the exchange of heavy neutral Dirac fermions implied by the
matrix
\begin{equation}
M_\nu=\left(
\begin{array}{ccc}
0&M_D&0\\
M_D^T & 0 &M\\
0&M^T& 0
\end{array}
\right),
\end{equation}
(in the basis $\nu,\,\nu^c, \, S$) will keep the neutrinos massless
and yet allow for LFV effects.
This is the idea behind the so--called inverse seesaw
model~\cite{mohapatra:1986bd,bernabeu:1987gr} (for other extended
seesaw schemes see,
e.g.~\cite{Wyler:1983dd,Akhmedov:1995vm,Barr:2005ss}).
Note that, to each of the isodoublet neutrinos $\nu$ two \321
isosinglets $\nu^c, \, S$ are added~\footnote{For simplicity we add
  the isosinglet pairs sequentially, though more economical variants
  may be possible.}.
Neutrinos get masses only when $U(1)_L$ is broken, for example through
a nonzero $\mu SS$ mass term. Thanks to the lepton number symmetry
which arises as $\mu\to 0$ the magnitude of $\mu$ can be chosen to be
small in a natural way, in the sense of 't
Hooft~\cite{'tHooft:1979bh}. Moreover, in specific models, the
smallness of $\mu$ may be dynamically
preferred~\cite{Bazzocchi:2009kc}.
After $U(1)_L$ breaking the effective light neutrino mass matrix
is given as
\begin{equation}\label{inv}
M_\nu=M_DM^{T^{-1}}\mu M^{-1}M_D^T.
\end{equation}
so that, when $\mu$ is small, $M_\nu$ is also small, even when $M$
lies at the electroweak scale.  In other words, the smallness of
neutrino masses does not require superheavy physics.

\section{Type-III seesaw variants}
\label{sec:type-iii-seesaw}

We now turn to simple variants of the above schemes where the \321
singlet fermions $\nu^c$ are replaced by SU(2) triplets
$\Sigma$~\cite{Foot:1988aq}.

\subsection{Normal type III seesaw }
\label{sec:normal-type-iii}

The minimal type III seesaw model is described by the Lagrangian
\begin{equation}
\mathcal{L}_{III}=M_{l_{ij}}L_il^c_jH
%+ i Tr (\overline\Sigma D{\hskip-2.5mm /} \Sigma)
+Y_{D_{ij}} L_i \Sigma_j \tilde{H}
-\frac{1}{2}M_{\Sigma_{ij}}\text{Tr}(\Sigma_i \Sigma_j) +\text{h.c.}
\end{equation}
where
\begin{equation}
\Sigma=\left(
\begin{array}{cc}
\Sigma^0/\sqrt{2}&\Sigma^+\\
\Sigma^-&-\Sigma^0/\sqrt{2}
\end{array}
\right)
\end{equation}
denotes the hyperchargeless isotriplet fermion, $Y(\Sigma)=0$ and
$H=(\phi^+,\phi^0)^T$ is the Standard Model Higgs scalar doublet.
The effective neutrino mass matrix is fully analogous to
Eq.~(\ref{normal}) and its smallness requires a very large isotriplet
fermion mass.

The charged lepton mass matrix is a $6\times 6$ matrix,
\begin{equation}
\label{eq:mlep}
M_{lep}=\left(
\begin{array}{ccc}
M_l&M_D\\
0 & M_{\Sigma}
\end{array}
\right)
\end{equation}
which is brought to diagonal form by a 6~$\times$~6 unitary matrix
$V_{\alpha \beta}$ of the same dimension, $\alpha, \beta=1,..,6$
$$
V^\dagger M_{lep}M_{lep}^\dagger V= (M_{lep}^{diag})^2, 
$$ 
leading to three light fermions, namely $e,\mu$ and $\tau$, and three
heavy charged fermions $C_i$ with $i=1,2,3$.

In analogy with the matrix describing neutrino NC interactions in
general type-I and type-II seesaw schemes introduced in
Ref.~\cite{schechter:1980gr} we define the $\mathcal{P}$ matrix as
below
\begin{equation}
\label{eq:gim-c}
\mathcal{P}=
\left(\begin{array}{ccc}
\mathcal{P}_{LL}&\mathcal{P}_{LH}\\
\mathcal{P}_{HL}&\mathcal{P}_{HH}
\end{array}\right).
\end{equation}
The piece
\begin{equation}\label{P} 
\mathcal{P}_{LL}=1-M_D^\dagger M_\Sigma^{-2}M_D 
\end{equation}
characterizes the NC Lagrangian of charged leptons in the mass basis
\begin{equation}\label{Pe}
\mathcal{L}_{NC}=\frac{g'}{c_W}{\mathcal{P}_{LL_{i\alpha}}}\overline{L}_i\gamma_\mu (g_V-g_A\gamma_5) L_\alpha \,
Z^{\mu}.
\end{equation}
For finite $M_\Sigma$ values there are non-diagonal elements
$\mathcal{P}_{LL}$ that induce tree level FCNC among the charged
leptons $e,\mu,\tau$. 
In other words the mixing between different isospins implies the
violation of the GIM mechanism~\cite{lee:1977ti} with amplitude of
order $\epsilon^2$ where $ \epsilon^2 \sim m_\nu M_\Sigma~$.
\begin{figure}
\begin{center}
\includegraphics[angle=0,height=5.5cm,width=0.45\textwidth]{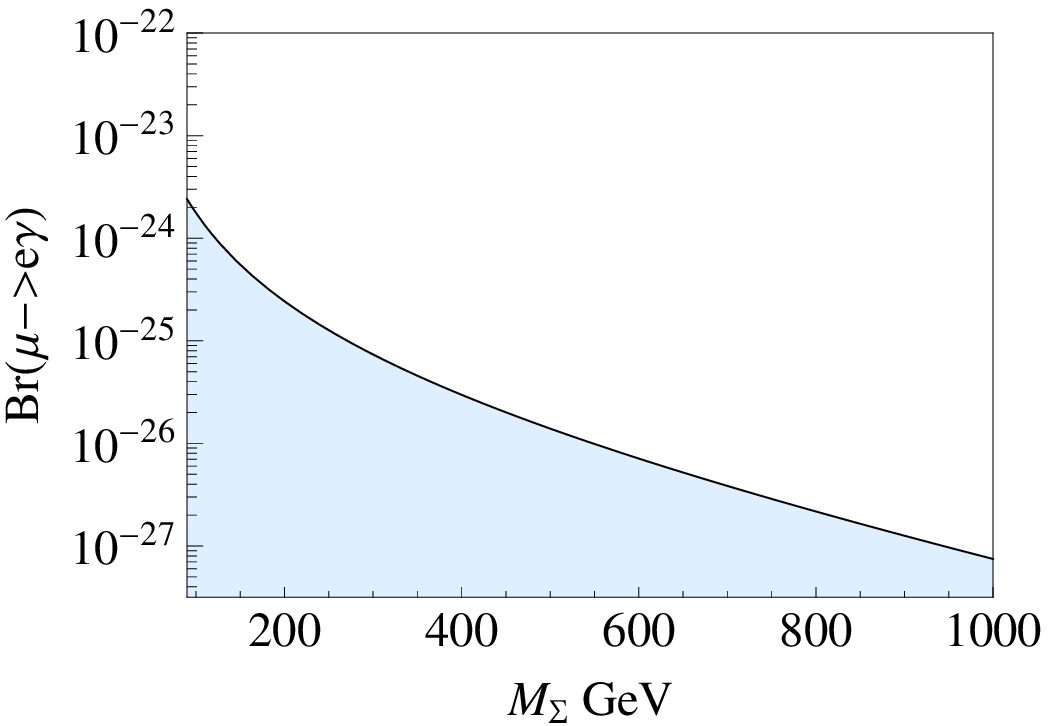}
\includegraphics[angle=0,height=5.5cm,width=0.45\textwidth]{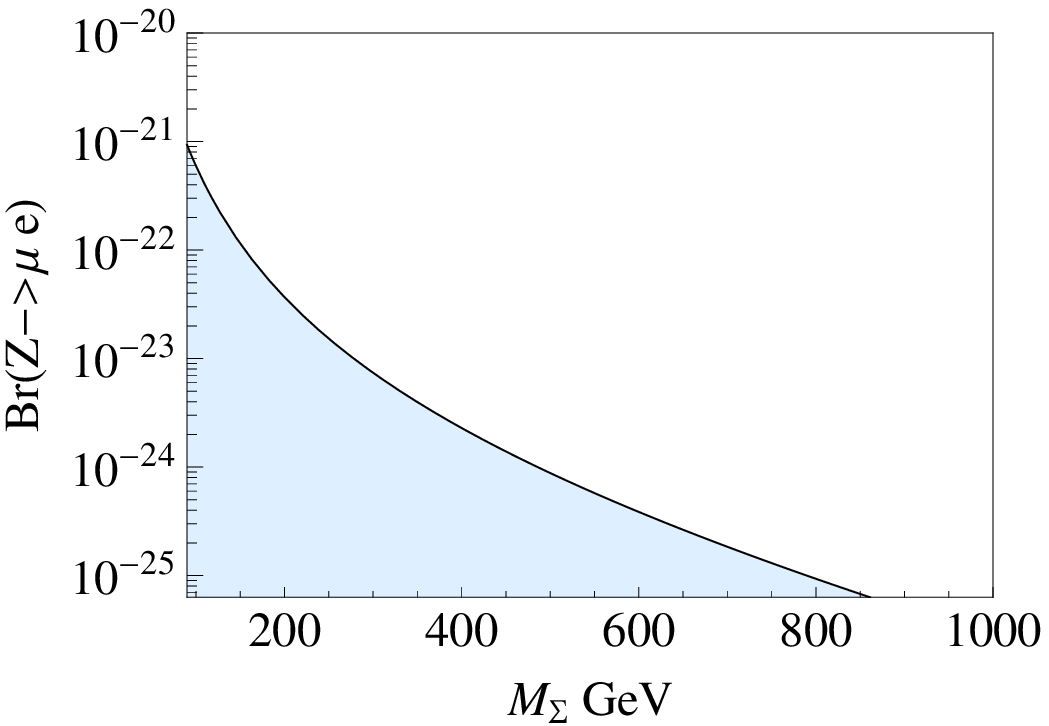}
\caption{Maximun attainable $\mu \to e \gamma$ and $Z\to e\mu$ decay
  branching ratios in normal type-III seesaw.}
\label{ref:fig1}
\end{center}
\end{figure}
The smallness of neutrino masses implies that, for $M_\Sigma$ values
accessible at the LHC, and barring fine-tuned parameter choices, the
expected LFV rates are expected to be too small to be of
phenomenological interest. We have estimated the maximum attainable
values for (i) the tree level LFV Z vertex, which also induces the
$l_i\to l_j \bar{l}_k l_m$ decays, and (ii) for the electromagnetic
penguin vertices, which induce the radiative LFV decays and $\mu-e$
conversion. Barring fine-tuning, one finds that they are far from the
sensitivities expected in the upcoming LFV
searches~\cite{Kuno:2000kd,vanderSchaaf:2003ti,Maki:2008zz}.
As an example Fig.~\ref{ref:fig1} illustrates the expected rates for
LFV Z-decay process.  Similarly LFV processes involving taus are too
small.

\subsection{Inverse type-III seesaw }
\label{sec:inverse-type-iii}

Having discarded normal type-III seesaw~\footnote{We consider here a
  non-supersymmetric model. Supersymmetry adds new sources of LFV.} as
an interesting model for lepton flavor violation, we turn instead to
an inverse type-III seesaw variant, characterized generically by the
Lagrangian

%%%%%%%%%%%%%%%%%%%%%%%%%%%%%%%%%%%%%%%%%%%%%%%%%%%%%%%%%%%%%

\begin{equation}
\begin{array}{l}
\mathcal{L}_{inv} = {Y_D}_{ij} \, L_i \, \Sigma_j \tilde{H}+
{Y_M}_{ij}\,  \text{Tr}(\Sigma_i \Delta ) S_j + \\
\qquad +\mu_{ij} S_{i} S_{j} + Y_{l_{ij}}   L_i\, l^c_j H~
-\frac{1}{2}M_{\Sigma_{ij}}\text{Tr}(\Sigma_i \Sigma_j),
\end{array}
\end{equation}
where, as before, $H=(\phi^+,\phi^0)^T$ denotes the Standard Model
Higgs scalar doublet and now $\Delta$ is a hyperchargeless scalar
$SU(2)$-triplet, $Y(\Delta)=0$
\begin{equation}
\Delta=\left(
\begin{array}{cc}
\Delta^0/\sqrt{2}&\Delta^+\\
\Delta^-&-\Delta^0/\sqrt{2}
\end{array}
\right),
\end{equation}
leading to 
\begin{equation}
\label{eq:minv}
M_\nu=\left(
\begin{array}{ccc}
0&M_D&0\\
M_D^T&M_\Sigma&M\\
0&M^T&\mu
\end{array}
\right).
\end{equation}
This leads to six heavy states $N_j$ with $j=4,..,9$ and an effective
three light Majorana eigenstates $\nu_{i}$ with $i=1,2,3$. The light
effective neutrino mass matrix is similar to that of the inverse
seesaw model with isosinglet instead of isotriplets~\footnote{We
  neglect loop contributions which exist due to the nonzero value of
  $M_\Sigma$. For an alternative inverse seesaw model with two lepton
  triplets see Ref.~\cite{Ma:2009kh}.}.
 \begin{equation}\label{inv2}
M_\nu\approx M_DM^{T^{-1}}\mu M^{-1}M_D^T.
\end{equation}
The smallness of the parameter $\mu$ may also arise
dynamically~\cite{Bazzocchi:2009kc} and/or spontaneously in a
Majoron-like scheme with $\mu\sim \vev{\sigma}$ where $\sigma$ is a
\321 singlet \cite{GonzalezGarcia:1988rw}. In the latter case, for
sufficiently low values of $\vev{\sigma}$ there may be Majoron
emission effects in neutrinoless double beta
decay~\cite{berezhiani:1992cd}.

Note that now the ratio $\epsilon\sim M_D M_\Sigma^{-1}$ need not be
too small to reproduce acceptably small neutrino masses, since the
latter vanish in the limit where the parameter $\mu$ goes to
zero~\cite{mohapatra:1986bd}. The smallness of $\mu$ is not only
natural~\cite{'tHooft:1979bh} but also dynamically preferred in some
cases~\cite{Bazzocchi:2009kc}.  The smaller the $\mu$ values the
larger can be the $\epsilon$. This implies that when the mass of
$\Sigma$ is accesible at LHC, say of the order of TeV, one expects
relatively large LFV decay rates.
In fact the situation is completely novel with respect to what one is
used to, in the sense that LFV as well as CP violation effects survive
even in the limit when neutrinos become
massless~\cite{bernabeu:1987gr} \cite{branco:1989bn}.
Clearly now FCNC effects can be naturally enhanced without conflict
with the smallness of neutrino masses.

\section{Tri-bimaximal inverse type-III seesaw }

The neutrino mixing angles~\cite{Maltoni:2004ei} indicated by neutrino
oscillation
experiments~\cite{fukuda:2002pe,ahmad:2002jz,araki:2004mb,Kajita:2004ga,ahn:2002up}
should be explained from first principles.
Here we consider the possibility of doing so in the framework of the
inverse \321 seesaw mechanism. To this end we adopt the attractive
tribimaximal (TBM) ansatz for lepton mixing~\cite{Harrison:2002er}
\begin{equation}
\label{eq:HPS}
U_{\textrm{HPS}} = 
\left(\begin{array}{ccc}
\sqrt{2/3} & 1/\sqrt{3} & 0\\
-1/\sqrt{6} & 1/\sqrt{3} & -1/\sqrt{2}\\
-1/\sqrt{6} & 1/\sqrt{3} & 1/\sqrt{2}
\end{array}\right)
\end{equation}
which provides a good first approximation to the values indicated by
current neutrino oscillation data.

Here we propose a simple $A_4$ flavor symmetry realization of the TBM
lepton mixing pattern within the inverse type-III seesaw scheme.  An
$A_4$ realization of the TBM in inverse seesaw has already been
studied in \cite{Hirsch:2009mx}.

Recall that $A_4$ is the group of the even permutations of four
objects.  Such a symmetry was introduced to yield
$\tan^2\theta_{\Atm}=1$ and $\sin^2\theta_{\textrm{Chooz}}=0$
\cite{Ma:2001dn,babu:2002dz,Hirsch:2003dr}.  Most recently $A_4$ has
also been used to derive $\tan^2\theta_{\Sol}=0.5$
\cite{Altarelli:2005yp}.  The group $A_4$ has 12 elements and is
isomorphic to the group of the symmetries of the tetrahedron, with
four irreducible representations, three distinct singlets $1$, $1'$
and $1''$ and one triplet $3$. For their multiplications see for
instance Ref.~\cite{Altarelli:2005yp}.
The matter fields are assigned  as in table \ref{tab2}.\\
\begin{table}[h!]
\begin{center}
\begin{tabular}{|c|c|c|c|c||c|c|c|}
\hline
&$L$ & $l^c$ & $\Sigma$& $S$ &  $\xi, \phi$& $\xi'\phi'$ & $\Delta$ \\
\hline
$SU_L(2)$ & 2& 1& 3& 1 & 2&2& 3\\
$Z_3$&$\omega$ &  $\omega$&$1$ &1 &   $\omega^2$&$\omega$ &1 \\
$A_4$&3&3&3&3&1,3&1,3 &1\\
\hline
\end{tabular}\caption{Matter assignment for inverse seesaw model.}\label{tab2}
\end{center}
\end{table}

The renormalizable~\footnote{ Here we have introduced several Higgs
  doublets. We can equivalently avoid having many Higgs doublets by
  introducing corresponding scalar electroweak singlet {\it flavon}
  fields.}  Lagrangian invariant under the $A_4\times Z_3$ symmetry is
\begin{equation}
\begin{array}{l}
\mathcal{L} = {Y_D}^k_{ij} \, L_i \, \Sigma_j \phi_k+{Y_D} \, L_i \, \Sigma_i \xi 
+{Y_M}_{ij}\,  \Sigma^0_i S_j \Delta+ \\
\mu_{ij} S_{i} S_{j} + Y_{l_{ij}}^k   L_i\, l^c_j \phi'_k+ Y_{l}   L_i\, l^c_i \xi' 
-\frac{1}{2}M_{\Sigma}\text{Tr}(\Sigma_i \Sigma_j)
\end{array}
\end{equation}
where from $A_4$-contractions one finds $\mu_{ij}\equiv \mu I_{ij}$,
$M_{ij}=M I_{ij}$. When $\xi$ takes a vacuum expectation value (vev)
and $\phi$ takes a vev along the $A_4$ direction
\begin{equation}\label{phi1}
\vev{\phi} \sim (1,0,0),
\end{equation}
we generate the Dirac mass entry, given as
\begin{equation}\label{eq:DA4}
M_D=\left(
\begin{array}{ccc}
a&0&0\\
0&a&b_1\\
0&b_2&a
\end{array}
\right),
\end{equation}
where we will also assume $b_1=b_2=b$. Such a relation can be obtained
in the context of $SO(10)$. 
Moreover, when $\xi'$ and $\phi'$ take on nonzero vevs, the latter along the
$A_4$ direction
\begin{equation}\label{phi2}
\langle \phi' \rangle\sim (1,1,1),
\end{equation}
we induce the charged lepton mass matrix as 
\begin{equation}
M_l=\left(
\begin{array}{ccc}
\alpha&\beta&\gamma\\
\gamma&\alpha&\beta\\
\beta&\gamma&\alpha
\end{array}
\right)=
U_\omega\left(
\begin{array}{ccc}
m_e&0&0\\
0&m_\nu & 0 \\
0 & 0&m_\tau
\end{array}
\right)U_\omega^\dagger.
\end{equation}
where 
$$
U_\omega=\frac{1}{\sqrt{3}}\left(
\begin{array}{ccc}
1&1&1\\
1&\omega & \omega^2 \\
1 & \omega^2&\omega
\end{array}
\right).
$$
The light neutrino mass matrix is diagonalized by 
\begin{equation}\label{DA4}
V_\nu=\left(
\begin{array}{ccc}
0 & 1 &0\\
1/\sqrt{2}&0&-i/\sqrt{2}\\
1/\sqrt{2}&0&i/\sqrt{2}
\end{array}
\right)
\end{equation}
and the corresponding eigenvalues are
\begin{equation}\label{eigiss}
\{m_1,m_2,m_3\}=\frac{v_\mu}{v_M^2}
\{(a+b)^2,a^2,-(a-b)^2\}.
\end{equation}
It follows that the lepton mixing matrix $U^\dagger_\omega\cdot V_\nu$
is the tri-bimaximal matrix.

As seen in Eq.~(\ref{eq:mlep}) the couplings $L\Sigma\phi,\,L\Sigma \xi$ give
us an off-diagonal block to the following $6$ by $6$ charged lepton
mass matrix for $L$ and $\Sigma$.
As a result the GIM mechamism is violated and there are FCNC among the
charged leptons $e,\mu,\tau$ at the tree level. 

Note that when the Higgs doublets $\phi$ and $\phi'$ take nonzero
vevs, the $A_4$ symmetry breaks spontaneously into its $Z_2$ and $Z_3$
subgroups, respectively. Such a {\it misalignment} implies a large
mixing in the neutrino sector. The implemention of such alignment has
been studied in many contexts
\cite{Altarelli:2005yx,Altarelli:2005yp,Ma:2008ym,Grimus:2008tt,
  Zee:2005ut,Hirsch:2008mg,Morisi:2009qa,Grimus:2008vg}.

\section{LFV in inverse type-III seesaw}

A characteristic feature of our seesaw scheme based on the use of
isotriplet instead of isosinglet lepton exchange is the existence of
tree level FCNC among the charged leptons.
While typically small in high-scale type-I seesaw, LFV effects are
well known to be potentially large in low-scale seesaw schemes, such
as the inverse~\cite{mohapatra:1986bd,Bazzocchi:2009kc} or the linear
seesaw~\cite{Malinsky:2005bi}.
In fact, in such schemes LFV rates are restricted only by weak
universality limits~\cite{bernabeu:1987gr}
\cite{gonzalez-garcia:1992be,Ilakovac:1994kj,Deppisch:2004fa,Deppisch:2005zm}
evading all constraints from the observed smallness of neutrino
masses.

We now consider an inverse seesaw scheme based on an underlying $A_4$
flavor symmetry. In contrast to Ref.~\cite{Hirsch:2009mx} we consider
now a type-III seesaw variant. For simplicity we neglect contributions
from Higgs boson exchange, which is a reasonable approximation.
We divide the LFV decay processes into three classes: A) $Z\to l_i
\bar l_j$, B) $l_i\to l_j \bar{l}_k l_m$ which proceed at the tree
level, and C) the loop-calculable $l_i\to l_j \gamma$ decays.

Note that in our model we have only two parameters in the Dirac mass
matrix plus a relative phase, and two extra TeV-scale parameters
$M,M_\Sigma$, in addition to the small parameter $\mu$ characterizing
the low-scale violation of lepton number. Two of these parameters are
determined by solar and atmospheric splittings~\cite{Maltoni:2004ei}.

Note also that the two parameters $M,M_\Sigma$ may be traded for the
heavy lepton mass $M_{N}$, and the mixing $\cos\theta_{\Sigma S}$
which will specify its production cross section at the LHC, through
the following rotation $(\Sigma_\alpha,S_\beta)$
\begin{equation}
\left(\begin{array}{cc}
\cos \theta_{\Sigma S}I&\sin \theta_{\Sigma S}I\\
-\sin \theta_{\Sigma S}I&\cos \theta_{\Sigma S}I
\end{array}\right)
\end{equation}
As we will see, the mass matrices are expressed in terms of very few
parameters, with a strong impact in the expected pattern of LFV
decays.

\subsection{$Z\to l_i \bar l_j$}

In our model the charged lepton mass matrix is a $6\times 6$ matrix,
which is brought to left-diagonal form by corresponding unitary matrix
$V_{\alpha \beta}$ of the same dimension, $\alpha, \beta=1,..,6$,
leading to three light fermion masses, namely $e,\mu$ and $\tau$, and
three heavy charged fermions $C_i$ with $i=1,2,3$.

Defining the $\mathcal{P}$ matrix as in Eq.~(\ref{eq:gim-c}) one
expresses the NC Lagrangian in the mass basis as in Eq.~(\ref{Pe})
where
\begin{equation}\label{PP}
\mathcal{P}_{LL}=1-U_\omega^\dagger M_D^\dagger M_\Sigma^{-2}M_D U_\omega
\end{equation}
This implies that for $i \ne j$ we have
\begin{equation}
\Gamma(Z\to l_i \bar l_j)=\frac{G_FM_Z^3}{6\sqrt{2}\pi} ({g_V^{l}+g_A^l})^2 
|{\mathcal{P}_{LL_{ij}}}|^2,
\end{equation}
where $g_A$ and $g_V$ are respectively the axial and vector couplings
of the charged leptons. This way one gets an effective
GIM-mechanism-violating vertex which possesses a well-defined
structure that follows from the flavor symmetry. This relates ratios
of branching ratios of FCNC decays. However, none of these decays is
allowed to be large in view of the stringent bounds on LFV muon
violating decays, see below.

\subsection{$l_i\to l_j \bar{l}_k l_m$}

This process occurs through the exchange of a virtual $Z$ boson, due
to the basic $Z l_i \bar l_j$ vertex.  The resulting branching ratio
is
$$
\Gamma(l_i\to l_j \bar{l}_k l_m)=\frac{G_Fm_i^5}{192\pi^3}Q_i Q_k |{\mathcal{P}_{LL_{ij}}}{\mathcal{P}_{LL_{km}}}|^2 
$$
where $Q_{i}$ are the electroweak charges defined as
$g_V^{l_i}+g_A^{l_i}$ for left-handed fields and as
$g_V^{l_i}-g_A^{l_i}$ for right-handed fields. Note that in contrast
to the case of the $Z$-decay which is proportional to $\epsilon^4$,
the three body decay with double LFV is proportional to $\epsilon^8$
and hence irrelevant. As we will show in Table~\ref{tab3} even the tau
decays that fo as $\epsilon^4$ will turn out to be small once the muon
decay constraints are implemented.

\begin{figure}[h]
\begin{center}
\includegraphics[angle=0,height=6cm,width=0.45\textwidth]{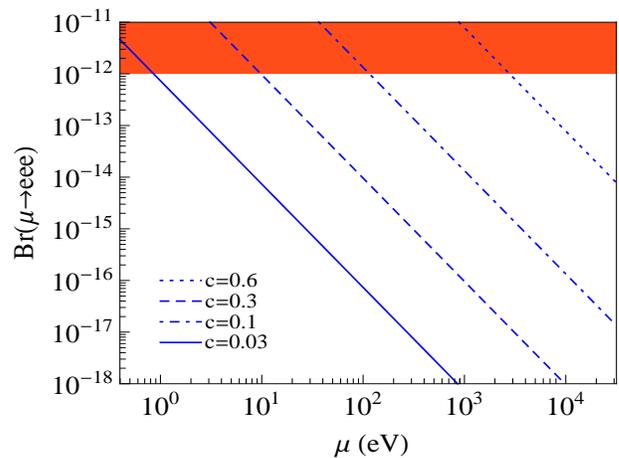}
\caption{ Branching of $\mu$ decay into $3e$ as a function of the
  $\mu$ parameter for different values of $c$ equivalent to 
  $\cos \theta_{\Sigma S}$,
  0.6 (dotted), 0.3 (dot-dashed), 0.1 (dashed) and 0.03 (contineous).
  Here $M_{N}$ is fixed at 1 TeV.}
\label{fig:3e}
\end{center}
\end{figure}
In Fig.~\ref{fig:3e} we present the dependence of the $\mu\to eee$
branching ratio on the $\mu$ parameter that characterizes the lepton
number violation scale, for a fixed value of the $M_{N}$. Although LFV
exists in the limit where neutrinos go massless, there is an indirect
dependence on the value of $\mu$ reflecting the need to account for
neutrino oscillation data.

\subsection{  $l_i \to l_j \gamma$}

\begin{figure}
\begin{center}
\includegraphics[angle=0,height=4cm,width=0.24\textwidth]{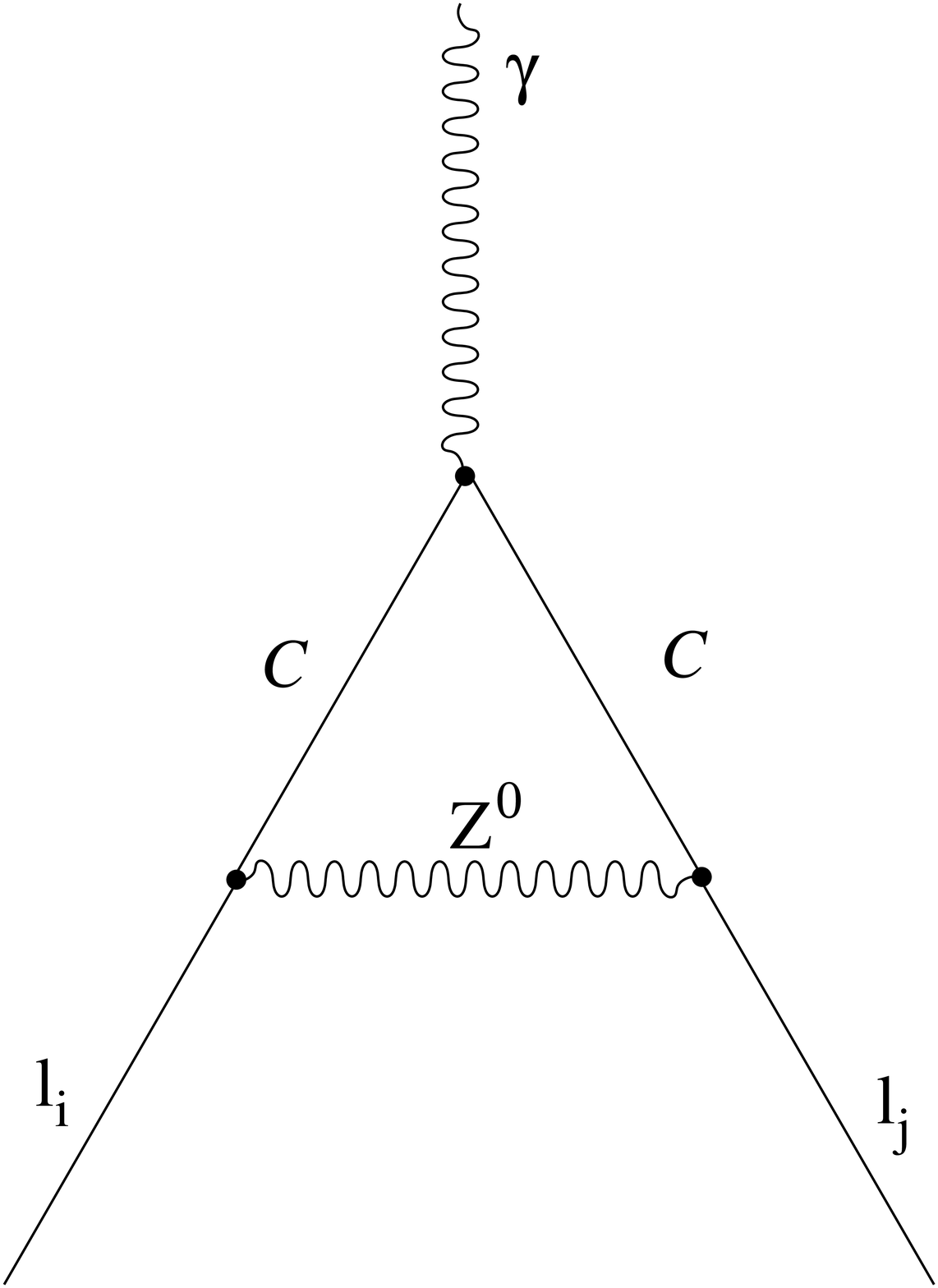}
\includegraphics[angle=0,height=4cm,width=0.24\textwidth]{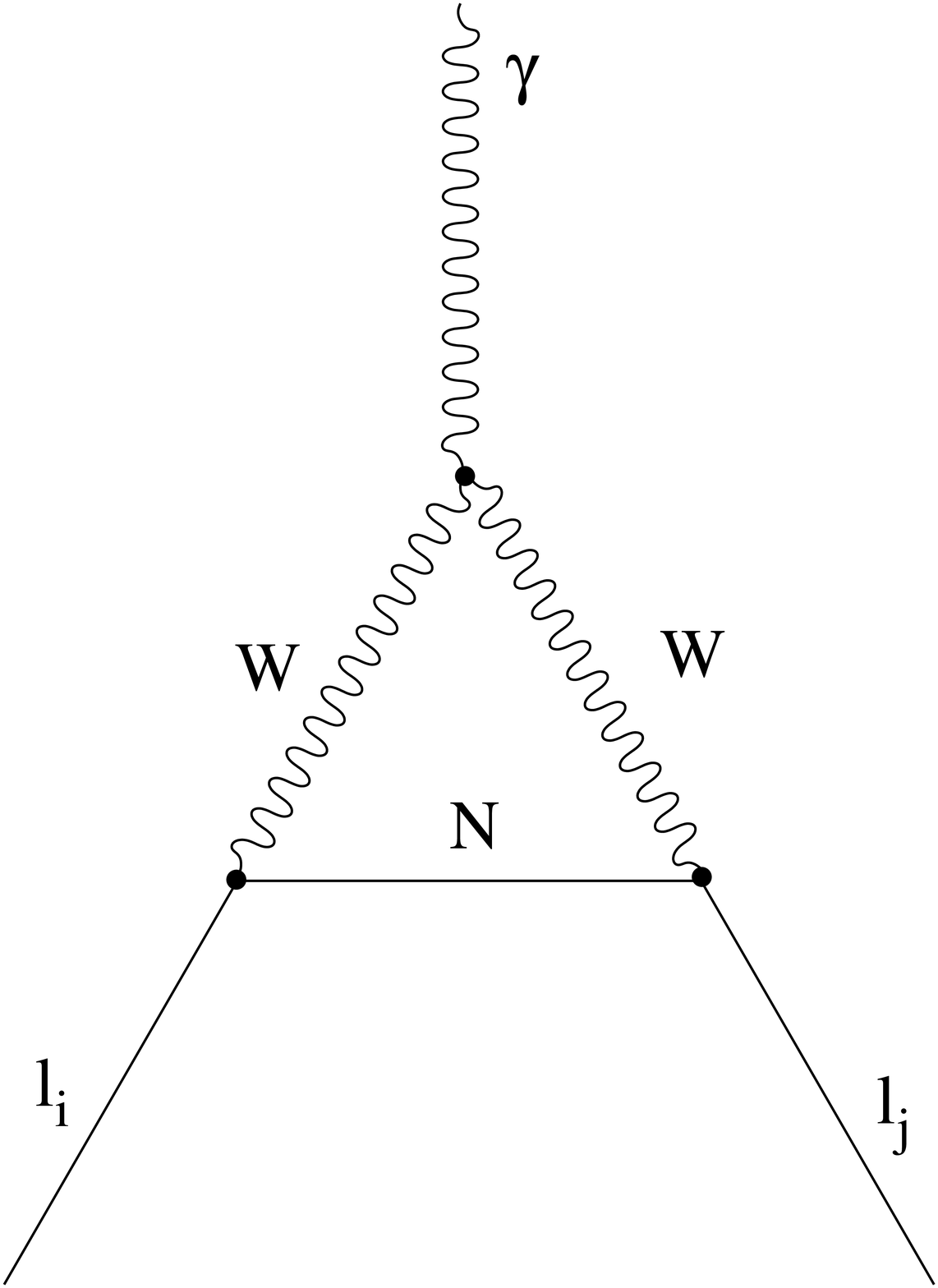}
\caption{Feynman graphs for $\mu \to e \gamma$ decay in type-III
  seesaw models.}
\label{ref:feynmangraphs}
\end{center}
\end{figure}

The decay \(l_i \to l_j \gamma\) arises in our model at one loop both
from charged as well as neutral current contributions, see
Fig.~\ref{ref:feynmangraphs}.
The neutrino mass matrix in Eq.~(\ref{eq:minv}) is a $9\times 9$
symmetric matrix, diagonalized by a unitary matrix $U_{\alpha\beta}$.
The effective charged current weak interaction is characterized by a
rectangular lepton mixing matrix
$K_{i\alpha}$~\cite{schechter:1980gr}
\begin{equation}
\mathcal{L}_{CC}=\frac{g}{\sqrt{2}}K_{i\alpha}\overline{L}_i\gamma_\mu (1+\gamma_5) N_\alpha \, W^{\mu}~, 
\end{equation}
where $i=1,2,3$ denote the left-handed charged leptons and $\alpha$
the label the neutral states, $\alpha,~\beta=1...9$.

Similarly the effective neutral current weak interaction of the
left-handed charged leptons with the heavy charged fermions is
characterized by
\begin{equation}
\mathcal{L}_{NC}=\frac{g}{\sqrt{2}}{\mathcal{P}_{LH}}_{i\alpha}\overline{L}_i\gamma_\mu (1+\gamma_5) C_\alpha \, 
Z^{\mu}. 
\end{equation}
where $i=1,2,3$  and $\alpha=4,5,6$.

The \(l_i \to l_j \gamma\) decays occur mainly through the exchange of
the six neutral heavy leptons $N_j$ subdominantly coupled to the
charged leptons
\cite{bernabeu:1987gr,Ilakovac:1994kj,Deppisch:2004fa} and that of
the three heavy charged fermion triplets which couple to the charged
leptons through the exchange of neutral $Z^0$ gauge boson (see, for
instance \cite{Lavoura:2003xp}).
\begin{figure}[b]
\begin{center}
\includegraphics[angle=0,height=6cm,width=0.45\textwidth]{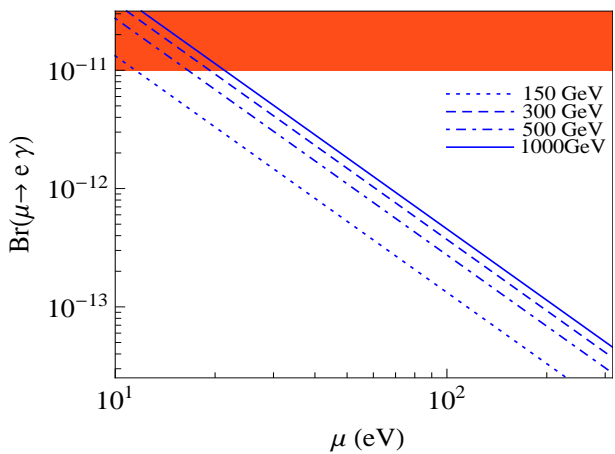}
\caption{
Branching of $\mu$ decay  into $e\gamma$ as a function of the $\mu$ parameter
for different values of $M_{N}$, $150$\,GeV (dotted), $300$\,GeV (dot-dashed),
$500$\,GeV (dashed) and $1$\,TeV (contineous) fixing $\cos \theta_{\Sigma S}=0.1$
}
\label{fig:meg}
\end{center}
\end{figure}

The resulting branching ratio is given by 
\begin{equation}
Br(l_i\to l_j \gamma)= \frac{\alpha^3s_W^2}{256 \pi^2}\frac{m_{l_i}^5}{M_W^4}
\frac{1}{\Gamma_{l_{i}}}|G_{ij}^W+G_{ij}^Z|^2
\end{equation}
where
\begin{equation}\label{def:G}
\begin{array}{l}
G_{ij}^W=\sum_{k=4}^9 K^*_{ik} K_{jk} G_\gamma^W\left(\frac{m^2_{N_k}}{M_W^2}\right)\\
G_{ij}^Z=\sum_{k=4}^6 {V^*_{ik}} {V_{jk}} G_\gamma^Z\left(\frac{m^2_{\Sigma_k}}{M_Z^2}\right)\\
G_\gamma^W(x)=-\frac{2 x^3+5 x^2 -x}{4 (1-x)^3}-\frac{3 x^3}{2(1-x)^4}\ln x\\
G_\gamma^Z(x)=-\frac{-5 x^3+9 x^2 -30 x +8}{(1-x)^3}-\frac{18 x^2}{(1-x)^3}\ln x.
\end{array}
\end{equation}

In Fig.~\ref{fig:meg} we study the dependence of the $\mu\to e\gamma$
decay branching ratio on the parameter $\mu$ which characterizes
lepton number violation. The same comment made in the discussion of
$\mu\to eee$ applies also here. Note that the branching $\mu\to
e\gamma$ depends somewhat on the physical mass $M_N$ of the neutral
heavy states, reflecting the fact that is a one loop process.

\subsection{Relating different LFV decays}

Note that, thanks to the admixture of the neutral and charged TeV
states in the weak interaction currents, the LFV branching ratios in
our inverse type-III seesaw model can be sizeable even in the absence
of supersymmetry.
Moreover, the assummed $A_4$ based flavor symmetry implies that the
structure of the matrices $K$ and ${\mathcal{P}}$ describing these
processes is special, leading to relationships among the LFV
branching ratios (see Table\,\ref{tab3} below).
As a result the $G^W,G^Z$ loop factor matrices of Eq.~(\ref{def:G})
and the $\mathcal{P}_{LL}$ matrix in Eq.~(\ref{Pe}) are determined by
just two model parameters,
\begin{eqnarray}
&&\quad G^W\sim G^Z\sim \mathcal{P}_{LL}\sim \\
&&\left(\begin{array}{ccc}
a^2 +\frac{4ab}{3}+\frac{2b^2}{3}& -\frac{1}{3}b(2a+b)&-\frac{1}{3}b(2a+b)\\
-\frac{1}{3}b(2a+b)  & a^2 -\frac{2ab}{3}+\frac{2b^2}{3} & \frac{1}{3}b(4a-b) \\
-\frac{1}{3}b(2a+b)&  \frac{1}{3}b(4a-b) & a^2 -\frac{2ab}{3}+\frac{2b^2}{3} 
\end{array}\right). \nn
\end{eqnarray}
Taking ratios of branching ratios, prefactors cancel and one finds for
example that
\begin{equation}\label{brin}
\frac{Br(\tau\to\mu\gamma)}{Br(\tau\to e \gamma)}=
\left(\frac{4+t}{2-t}\right)^2,
\end{equation}
where $t\equiv -b/a$ is the solution of the eq.
\begin{equation}\label{altin}
\alpha=\frac{1-(1-t)^4}{(1+t)^4-1}. 
\end{equation}
where the ratio $$\alpha= \Dms/\Dma$$ is well determined by neutrino
oscillation data~\cite{Maltoni:2004ei}.

The symmetry predictions are listed in Table~\ref{tab3}. They show
that, as long as the flavor symmetry holds, all LFV decay braching
ratios can be expressed in terms of the branching ratios for the
processes $\mu^- \to e^-e^+e^-$ and $\mu \to e \gamma$. 
Thanks to the tree-level violation of lepton flavor in the neutral
current, the relative ratio between $\mu^- \to e^-e^+e^-$ and $\mu \to
e \gamma$ is also unusual, and allows the rate for $\mu^- \to
e^-e^+e^-$ to be larger than that for $\mu \to e \gamma$.

Regarding the rates for mu-e conversion in nuclei, as already noted in
Ref.~\cite{Deppisch:2005zm}, in the limit where we neglect Higgs boson
contributions, these rates are strongly correlated with \(\mu \to e
\gamma\). This means that for a given target nucleus they are
relatively well determined from the $\mu \to e \gamma$ rate. We refer
the reader to Fig.~5 in Ref.~\cite{Deppisch:2005zm}.
Finally, tau LFV decay rates are expected to be small, even those that
scale as $\epsilon^4$ like those corresponding to semi-leptonic modes,
which are not displayed in the Table.
\begin{center}
\begin{table}[t!]
\begin{tabular}{|l|c|}
\hline
$\frac{Br(\mu^- \to  e^-e^+e^-)}{Br(\tau^- \to  e^-e^+e^-)}$&$\left(\frac{m_\mu}{m_\tau}\right)^5\frac{\Gamma(\tau \to  all)}{\Gamma(\mu \to  all)}$\\
\hline
$\frac{Br(\mu^- \to  e^-e^+e^-)}{Br(\tau^- \to  e^-\mu^+\mu^-)}$&$
\left(\frac{m_\mu}{m_\tau}\right)^5\frac{\Gamma(\tau \to  all)}{\Gamma(\mu \to  all)}$\\
\hline
$\frac{Br(\mu^- \to  e^-e^+e^-)}{Br(\tau^- \to  \mu^-e^+e^-)}$&$
\left(\frac{m_\mu}{m_\tau}\right)^5\frac{\Gamma(\tau \to  all)}{\Gamma(\mu \to  all)}\left(\frac{2-t}{4+t}\right)^2$\\
\hline
$\frac{Br(\mu^- \to  e^-e^+e^-)}{Br(\tau^- \to  \mu^-\mu^+\mu^-)}$&$
\left(\frac{m_\mu}{m_\tau}\right)^5\frac{\Gamma(\tau \to  all)}{\Gamma(\mu \to  all)}\left(\frac{2-t}{4+t}\right)^2$\\
\hline
$\frac{Br(\tau^- \to  \mu^-\mu^-e^+)}{Br(\tau^- \to  e^-e^-\mu^+)}$&$\left(\frac{4+t}{2-t}\right)^2$\\
\hline
\hline
$\frac{Br(Z^0 \to \mu^-e^+)}{Br(Z^0 \to \tau^-e^+)}$&1\\
\hline
$\frac{Br(Z^0 \to \mu^-e^+)}{Br(Z^0 \to \tau^-\mu^+)}$&$\left(\frac{2-t}{4+t}\right)^2$\\
\hline
\hline
$\frac{Br(\mu \to  e\gamma)}{Br(\tau \to  e\gamma)}$&$\left(\frac{m_\mu}{m_\tau}\right)^5\frac{\Gamma(\tau \to  all)}{\Gamma(\mu \to  all)}$\\
\hline
$\frac{Br(\tau \to  \mu\gamma)}{Br(\tau \to e\gamma)}$&$\left(\frac{4+t}{2-t}\right)^2$\\
\hline
\end{tabular}\caption{Predictions for ratio of LFV branching, where $t$ is defined in the text
and $(m_\mu/m_\tau)^5\Gamma_\tau/\Gamma_\mu=0.18$.
}\label{tab3}
\end{table}
\end{center}

\begin{figure}[h]
\begin{center}
\includegraphics[angle=0,height=6cm,width=0.45\textwidth]{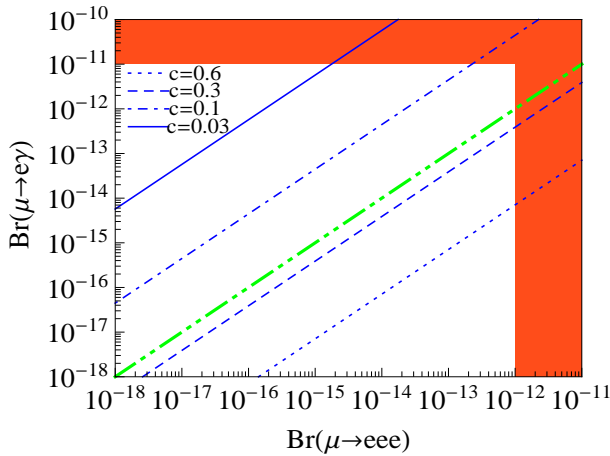}
\caption{Branching of $\mu$ decay  into $e\gamma$ {\it vs} the branching of $\mu$ decay into
$3e$ for different values of $\cos \theta_{\Sigma S}$, 0.6 (dotted), 0.3 (dot-dashed),
0.1 (dashed) and 0.03 (continous) and $M_{N}=1$~TeV.}
\label{fig:e3vseg}
\end{center}
\end{figure}

\section*{Discussion}

We have proposed a new inverted version of type-III seesaw or
equivalently, a new type-III version of the inverse seesaw mechanism.
This way the physics responsible for neutrino masses can lie at
low-scale and can be accessible at the Large Hadron Collider (LHC),
due to: (i) the TeV--scale neutral fermions having large cross
sections at the LHC and (ii) the TeV--scale neutral fermions inducing
large LFV processes due to the low-scale violation of the
Glashow-Iliopoulos-Maiani mechanism, which implies potentially large
tree-level FCNC involving charged leptons.
By assuming an $A_4$-based underlying flavor symmetry we have
implemented a tri-bimaximal lepton mixing pattern to account for the
observed neutrino oscillation parameters.
We have studied the phenomenology of the resulting LFV decays and
given the typical expectations for their magnitude, in addition to
discussing the predictions for their relative rates. 
In Fig.\,\ref{fig:e3vseg} we give the correlation between the
branching of $\mu \to e \gamma $ vs the branching of $\mu \to eee$
fixing $M_{N}$ and for different values of $\cos \theta_{\Sigma
  S}$. %% and $\mu$.  
Clearly neutral heavy fermion states can lie at the TeV scale and
their production cross section at the LHC is enhanced with respect to
that expected in type-I inverse seesaw~\cite{Dittmar:1990yg}.
Indeed the much larger production cross sections expected for the
type-III models should encourage detailed dedicated MonteCarlo
simulations~\cite{Franceschini:2008pz} in order to scrutinize the
viability of detecting the associated signals. Last but not least,
given the underlying flavor symmetry predictions these should also
take into account the details of flavor physics which will determine
the expected decay pattern of the heavy leptons.

\section*{Acknowledgments}

We thank Martin Hirsch and Ioannis Papavassiliou for useful
comments. This work was supported by the Spanish grant
FPA2008-00319/FPA and PROMETEO/2009/091.

%  \bibliographystyle{h-physrev4} 
%  \bibliography{valle-ref,bibt3,morisi-ref}%%,soko,snova,nu-rev06,parke-ref}

\end{document}